\documentclass{article}

\usepackage{PRIMEarxiv}

\usepackage[utf8]{inputenc} 
\usepackage[T1]{fontenc}    
\usepackage{hyperref}       
\usepackage{url}            
\usepackage{booktabs}       
\usepackage{amsfonts}       
\usepackage{nicefrac}       
\usepackage{microtype}      
\usepackage{lipsum}
\usepackage{fancyhdr}       
\usepackage{graphicx}       
\usepackage{booktabs}
\graphicspath{{media/}}     

\newcommand{\paragraphNoSpace}[1] {\noindent\textbf{#1.~}}
\newcommand{\paragraphWithSpace}[1]{\vspace{1.25mm}\noindent\textbf{#1.~}}

\title{Democratization of Retail Trading: Can Reddit’s WallStreetBets Outperform Investment Bank Analysts?
}

\author{
  Tolga Buz \\
  Hasso Plattner Institute \\
  Potsdam, Germany\\
  \texttt{tolga.buz@hpi.de} \\
   \And
  Gerard de Melo \\
  Hasso Plattner Institute \\
  Potsdam, Germany\\
  \texttt{gdm@demelo.org} \\
}

\begin{document}
\maketitle

\begin{abstract}
  The recent hype around Reddit's WallStreetBets (WSB) community has inspired research on its impact on our economy and society.
  Still, one important question remains:
  Can WSB's community of anonymous contributors actually provide valuable investment advice and possibly even outperform top financial institutions?
  We present a data-driven empirical study of investment recommendations of WSB in comparison to recommendations made by leading investment banks, based on more than 1.6 million WSB posts published since 2018. 
  To this end, we extract and evaluate investment recommendations from WSB's raw text for all S\&P 500 stocks and compare their performance to more than 16,000 analyst recommendations from the largest investment banks.
  While not all WSB recommendations prove profitable, our results show that they achieve average returns that compete with the best banks and outperform them in certain cases.
  Furthermore, the WSB community has been better than almost all investment banks at detecting top-performing stocks.
  We conclude that WSB may indeed constitute a freely accessible, valuable source of investment advice.
\end{abstract}

\keywords{Web mining \and Collective intelligence \and Collaborative investing \and Retail trading}

\section{Introduction}

The WallStreetBets (WSB) online community, one of many on the social media platform Reddit, skyrocketed in public awareness in early 2021, when it played a major role in the hype surrounding the GameStop stock and the resulting \emph{short squeeze}.
A brief look at popular posts on WSB typically reveals a large number of stock market-related memes and posts, in which community members envy each others' gains or ridicule their losses, all accompanied by vulgar language. At first glance, this may not appear to be a promising source for valuable investment advice, but rather one to visit for mere entertainment purposes. 
A closer look at the community, however, reveals that analysis reports (Due Diligences) authored by members and discussions about possible investment opportunities are an important part of WSB as well.
We seek to investigate whether an openly accessible, anonymous social community such as WSB can serve as a valuable source for stock market analysis and investment advice, especially when compared to trading recommendations from expert sources such as investment banks.

Reddit communities like WallStreetBets provide an anonymous and democratic place for millions of participants to share content or comment and vote on others' posts -- a post's success and visibility is thus decided by the community.
In contrast, most alternative sources for investment advice have in common that they tend to have a single source in control of what recommendations are published, and many of them either have to be paid or pursue indirect financial goals by advertising their own or others' paid products:
Investment banks release paid analyst reports; widely-known online news outlets such as The Motley Fool or Seeking Alpha provide paid access to their analyses (while publishing freely accessible articles that each recommend a subscription of their paid services); influential individuals share advice on their blogs or on social media platforms such as Twitter.
Hence, there is typically either a cost barrier to access investment advice or a certain degree of opacity regarding the author's motives and own investment activities.
For example, the publishers of The Motley Fool could exploit the impact and reach of their free publications to influence demand for stocks that they have recommended to their paying subscribers beforehand (this is of course hypothetical, as a separate in-depth analysis would be required to confirm or deny this).
Alternative forums on similar topics that we discovered during our research were difficult to find, had a very limited number of members (the largest ones with around 100,000 -- 300,000 users, in contrast to WSB's more than 12 million), and a limited amount of community activity.
Furthermore, the more professional ones among these forums are restricted by an application process or subscription fee.
Analyst reports from investment banks, which one might presume are the most proficient source of trading advice, remain difficult to access due to their prohibitively high prices \cite{brush2017research}.
If WallStreetBets can not only offer valuable investment advice, but compete with or even outperform the analysts of the world's largest investment banks, the community could thus be viewed as democratizing and facilitating access to valuable investment recommendations.

The WSB community shares a large number of posts every day, from which signals first need to be extracted, filtered, and evaluated correctly, in our case using data mining, engineering, and analysis techniques.
In addition, with regard to stock price movements, these signals may either be proactive (preceding them) or reactive (following them), which is a substantial difference in terms of their value for investors.
The months since the GameStop hype have sparked increased interest in WSB and a surge in what have since come to be known as ``meme stocks'', which are also the community's most actively discussed stocks.
Most meme stocks correspond to smaller companies with high growth potential, but high risk -- a more typical private investor or retail trader might consider them too risky and rather opt to invest in larger, more established companies.
The discourse on meme stocks is often accompanied by a multitude of posts about high profits as well as losses made by WSB users trading these stocks, leading to an increase in reactive discussions. 
Additionally, meme stocks often exhibit highly increased price volatility, which makes it much more difficult to time an investment correctly.
Although many meme stocks have performed well in the recent past, our analysis thus focuses on the S\&P 500 index of larger, more established companies in order to reduce effects caused by smaller meme stock companies as well as to create a level playing field for a comparison against more established sources.

Initial research we have conducted suggests that WSB may indeed be a source for investment advice \cite{buz2021should}.
The goal of this work is to examine the performance of WSB buy recommendations since 2018 on all S\&P 500 companies and compare them to the recommendations made by investment banks.
In particular, this includes a comparison of WSB against the top 20 investment banks, chosen based on the amount of analyst reports they have published on S\&P 500 companies within the same time frame.
Our results suggest that WSB's recommendations can keep up with the best professional investment banks, in some cases significantly outperforming them in terms of the average returns.

In this work, we present an approach of successfully extracting valuable information from the large amount of unstructured text posted on WallStreetBets.
Our research contributes to understanding the role and value of communities like WSB and shows that the community can serve as a valuable source for investment advice and offer an at least similarly attractive alternative to mainstream channels when signals are extracted with an appropriate methodology.

\section{Background}

\subsection{Reddit and WallStreetBets}

Reddit is an online social media platform that provides a place for communities (known as \emph{subreddits}) focusing on specific topics, such as humorous memes, politics, relationship advice, particular sports teams, or computer games, among numerous others.
Reddit was established in 2005 by Steve Huffman and Alexis Ohanian, and has since grown to host over 100,000 active subreddits with more than 52 million daily active participants, accumulating more than 50 billion monthly views \cite{redditinc}.
Members can post submissions (e.g., texts, links, images, videos), comment on them, use the up- or down-vote buttons to rate submissions and comments, and even gift authors with awards purchased beforehand. 

The subreddit WallStreetBets was created on January 31, 2012\footnote{\url{http://www.reddit.com/r/wallstreetbets}}.
On January 1, 2019 the number of subscribers was around 450,000, and by January 1, 2021, this figure had grown to around 1,760,000. Then, within just January 2021, the community witnessed a sudden staggering growth from 1.7 million to around 8.5 million \cite{wsbstats}, due to the media attention stemming from the GameStop hype. At the time of writing, the subreddit has more than 12 million subscribers.
The website subredditstats.com places WallStreetBets on rank 49 out of all subreddits regarding the subscriber count, and on rank 10 with respect to the amount of comments per day, 
indicating a highly active community. 
The subreddit describes itself as ``a community for making money and being amused while doing it. Or, realistically, a place to come and upvote memes when your portfolio is down.'' 
In comparison to other finance-related Reddit communities, WSB has the highest subscriber count and is notorious for its unique slang and the pervasive use of offensive terms. 
For instance, members of the community are officially referred to as ``degenerates'' and often refer to each other as ``apes'' or ``retards'' \cite{agrawal2022wallstreetbets}.

The unprecedented rise in popularity and news coverage in January 2021 mentioned above occurred after the community focused discussions on a series of stocks, including GameStop, that were in part deemed undervalued while simultaneously exhibiting a high short interest, i.e., the ratio of shares being sold short (or \emph{shorted}) by financial institutions.
Short-selling a stock refers to the practice of borrowing shares of a stock in order to sell them immediately with the goal of buying them back later at a lower price. 
This practice is a speculative move often employed by professional investors on stocks that are considered overvalued and expected to lose value in the near future.
However, if the stock price instead increases, a short position can lead to devastating losses.

While discussions of potentially undervalued investment opportunities with growth potential are common in investment-focused online communities, the fact that the GameStop stock showed a pronounced short interest by institutional investors led to the situation being portrayed as an ideological David-and-Goliath battle of small retail investors rallying against hedge funds: by buying and holding the stock, its price could be driven up, thus forcing the financial institutions to close their short positions at a significant loss. This, in turn, drove the stock price up even further, a phenomenon known as a \emph{short squeeze}, at which point the retail investors could sell their shares at a large profit.

These developments turned WallStreetBets into a place where the broader community of Reddit users could come together to unite in a movement, driven by the prospect of large financial gains through risky investments, with the added appeal of supposedly advancing the greater cause of punishing financial institutions, particularly hedge funds. The latter have been accused of ruining many people's lives during the financial mortgage crisis of 2008, among other events. In this paper, we assess more broadly whether the advice disseminated on WSB can keep up with that of professional bank analysts.

\subsection{Related Research}

Social media is widely considered a promising source of data to monitor the public sentiment on topics such as politics, companies, brands, and products, and is frequently studied especially in behavioural finance and decision sciences.
Research suggests that there are important ties between social media sentiment and macroeconomic factors such as consumer confidence \cite{daas2014social}, and that the general social media sentiment may affect a company's stock performance \cite{yu2013impact}.
Extensive research has attempted to show how particular cues from social media can enable predicting stock price changes \cite{sul2017trading, duz2020social, nguyen2015sentiment} -- however, most of these studies consider massive-scale aggregated data from Twitter rather than a close-knit community.

Another line of work focuses on the forms of interaction in modern online platforms. 
For instance, it has been shown that emotionally charged messages in social media can spread information more quickly \cite{stieglitz2013emotions}. Messages disseminated on social media, in some circumstances driven by bots, have been known to empower social movements \cite{manikonda2018twitter} and to have political influence on an international scale \cite{howard2011opening, gorodnichenko2018social}. 
One study investigated the social interaction on WallStreetBets \cite{boylston2021wallstreetbets}, explaining the nature of the community's conversations and language as of 2020.

Fuelled by the 2021 GameStop hype on WallStreetBets, recent research has provided a number of insights on the dynamics and background of the matter.
This includes studies on the social dynamics within the WSB community that led to the hype \cite{lyocsa2021yolo, semenova2021reddit} and on the idea of retail investors fighting against Wall Street \cite{chohan2021counter, di2021gamestop}. 
Other studies focused on the financial mechanisms driving the sudden price spike \cite{aharon2021did, anand2021wallstreetbets, wang2021predicting} and on the effect of retail traders on prices and volatility, along with their participation in transactions \cite{van2021equity, hasso2021participated, eaton2021zero}. 
Further research considered the implications of the events for market regulators and brokerages \cite{umar2021tale, macey2021securities, jones2021brokerages, feinstein2021clearing}. Yet another study provided an analysis of selected posts from an anthropological perspective \cite{mendoza2021sticking}.

The variety of related research shows that WallStreetBets and the GameStop hype can be investigated from multiple angles.
However, most recent research focuses on socio-economic and general market effects and implications, or very specific matters such as predicting price movements of a single stock using posts and comments. 
Little is known about the financial skills of WSB traders and the merits of their investment advice \cite{buz2021should} 
and even less about how their skills compare to those of alternative sources for investment advice.
Additionally, there is a lack of longitudinal research from an information science perspective -- studying data that span a longer time frame and a large number of different postings and assessing longer-term effects and trends.
This paper presents new insights from this data-driven perspective on the WSB community's proficiency in order to evaluate how well investors following the investment recommendations of WSB would have performed financially against those following the analyst reports of the largest investment banks.

\section{Methodology}

In order to have a sufficiently large basis for the analysis, we compiled an extensive dataset consisting of posts from WallStreetBets, along with financial information of all reviewed stocks obtained via the Yahoo!\ Finance API.
As a baseline portfolio, we consider the entirety of S\&P 500 stocks, which are widely viewed as representing the broader, established market.

On top of the collected data, we have established a methodology for information extraction and evaluation in order to achieve a standardized and comparable analysis across the large number of considered stocks.

\subsection{Data Acquisition from WallStreetBets}

\paragraphNoSpace{Data Compilation}
For our study, we review 1,614,976 WSB submissions in total, ranging from January 2018 to March 2022, using the Pushshift service \cite{baumgartner2020pushshift}.
In order to increase the quality of the data, we exclude all posts that have been deleted or removed (either by their author or moderators), as these posts are also not visible to WSB's visitors.
Our previous analysis \cite{buz2021should} has shown that it is beneficial for the quality of detected investment signals to utilize the community's post flair (label) system to select posts that are actually intended to provide predictive value -- these include discussions and investment analyses, but exclude memes and so-called ``shitposts'' among others.
This results in a cleaned dataset of 222,301 submissions.

It should be noted that a normal Reddit and WallStreetBets visitor would not be able to access the entirety of posts that are analyzed in this dataset during a single visit -- instead, the standard post ranking (``hot'' posts) are the most successful posts regarding their upvote score of the last few days.
If users would like to see other posts, they can choose to view the newest ones, which shows a few hundred items that go back multiple days, or the top posts (in terms of upvote scores) of a specific time window that the user can choose, e.g., a day, month, or year.
In contrast, our analysis assumes a normalized or equal treatment of all relevant posts as long as they are not removed or deleted, in order to be as representative as possible.
This amount of exposure simulates a user that visit the community at least two to three times per week to remain up-to-date and observe all relevant posts.
We consider this amount of activity realistic and assume that most active WSB users visit the community even more frequently.

\paragraphWithSpace{Attributes}
Nearly all submissions have a categorical label (referred to as ``flair'' in Reddit), which in the case of WSB distinguishes between discussions, memes, news, etc.
The distribution of submission flairs shows that the WSB community enjoys serious subject discussions, news, and analyses to a similar extent as posts for entertainment purposes, such as memes, gain and loss posts -- the flairs in our dataset are distributed as follows: 
Discussion (121,228), YOLO\footnote{YOLOs are high risk and high value investments usually on a single stock.} (32,805), DD\footnote{DD stands for Due Diligence, which generally consists of a detailed analysis of a stock or industry along with an explanation of why its value is likely to increase or decrease in the future.} (29,762), News (16,492), Options (5,578), Stocks (4,441), Technical Analysis (4,346), Fundamentals (2,340), Chart (1,985), Technicals (1,451), Daily Discussion (1,348), and Futures (525).
This excludes submissions labelled as Meme, Gain, Loss, Shitpost, Satire, Storytime, and Donation, which are all intended to be of reactive nature. For instance, Gain or Loss posts are made after an investment has been sold after holding it for some time, while Meme posts react to events in the community or in the stock market.

The flairs help in dividing submissions into two categories: those that are posted proactively in anticipation of stock price movements, e.g., DDs, YOLOs, Discussions, Options, and Technical Analysis, and those that show a person's reaction to stock price movement, e.g., Gain, Loss, and Meme.
In order to assess the predictive, proactive capabilities of the WSB community, we retain only the submissions for the former set of categories.
A comparison of results with the unfiltered dataset shows that focusing on ``proactive'' flairs consistently improves the quality of buy signals, confirming the assumption.

While the Pushshift service provides a variety of attributes along with each submission, many of them are Reddit-specific metadata that do not provide any informative signal for the purposes of our analysis. Hence, we focus on the data of each submission's title, body text (also called ``selftext'' in the Reddit API), timestamp, flair, and score. 
By combining the titles and body texts, we obtain one text per submission.
During this process, we tokenize the text snippets and discard punctuation and repetitive or unwanted symbols and characters.

\subsection{Selection and Detection of Stock Tickers}

\paragraphNoSpace{Reference Portfolio}
As the reference portfolio for comparing investment recommendations, we consider the companies of the S\&P 500 index, due to its popularity and widespread use among financial professionals as a representative index for the broader market.
The S\&P 500 covers the largest US companies listed on the stock market, which are distributed across many different industries.
In order to become part of the index, a company's stock must satisfy multiple criteria regarding market capitalization, amount of traded shares, liquidity, and earnings.
The companies included in the index are weighted by their market capitalization, currently leading to large technology companies such as Apple, Microsoft, and Google being the most important components of the index.

For the remainder of this study, we thus limit our focus to the list of all companies included in the index (as of March 2022).
It has to be noted that this approach may put WSB at a disadvantage compared to other investors, as the WSB community frequently engages in discussion of stocks of smaller companies with higher growth potential that are not listed on the S\&P 500. 
In particular, restricting the scope of the study excludes many of the meme stocks of WSB. 
On the plus side, this leads to a baseline portfolio that has shown less volatility and potential ``pump and dump'' activity (stocks that are hyped until a very high increase in price, just before they crash down to a price similar to or slightly higher than before).

For our analysis, we obtain market data on all S\&P 500 stocks, encompassing general stock information, the daily price history including open, close, high, low, and trading volume. This data is obtained via the Yahoo!\ Finance service.

\paragraphWithSpace{Detection of Stock Tickers on WSB}
For the analysis, we developed a detailed approach to detect stock mentions in WSB submissions.
For this, we iterate through all post submissions and detect those that contain a particular stock ticker.
In each such submission, considering the title and body text, we count the occurrences of a given ticker (in upper case, with and without a prefixed ``\$'', e.g. ``AAPL'' and ``\$AAPL'' for Apple, Inc.). 
For single-character tickers, such as Ford Motor Company's ``\$F'', we consider only occurrences with the dollar sign in order to avoid false positives, as our analysis has shown that single-character stock tickers are almost always written with the prefixed ``\$'' by the WSB community in order to avoid confusion, while occurrences without the prefix often do not refer to the stock: For example, the letter ``F'' is often invoked as an abbreviation for a popular swear word.
Additionally, we compile a list of S\&P 500 stock tickers that are often used as words or abbreviations with a meaning unrelated to the company, e.g., ``ALL'', ``IT'', ``LOW'', ``NOW'', ``ARE'', which we as well consider as mentions only when prefixed with ``\$''.

Having detected the relevant submissions for each of the stock tickers in the S\&P 500 index, we use two approaches to extract investment recommendations.
Our default approach computes the score of a ticker $t$ as \[f(t)=\sum_{w \in W_{+}} f(w,t)\, - \sum_{w \in W_{-}}f(w,t),\] i.e., it counts the (case-standardized) frequency $f(w,t)$ of buy-related words $w$ from a set $W_{+}$ (including ``buy'', ``call'', ``calls'', where calls refer to stock options that anticipate a stock price increase), from which it subtracts the counts of negation phrases $w \in W_{-}$ such as ``not buy'' and ``don't buy''.
Similarly, we apply this approach to ``hold'' and ``sell'' (including ``put'' / ``puts'' for the corresponding stock option).
We focus on words written in present tense, as past tense indicates that the post has been written retroactively.
In order to decide whether a single post can be deemed a buy, hold, or sell signal, we identify the highest of the three calculated values.
As a more sophisticated alternative to this method, we also consider another variant (denoted as ``WSB (prox)'' later on). 
In this case, $f(w,t)$ only considers occurrences of the  keywords occurring in close proximity (within 20 characters) of the stock ticker $t$.
After detecting which submissions provide an investment recommendation for a stock, we aggregate the number of recommendations per type for each day, resulting in a dataset that enables an analysis of WSB investment recommendations with a daily granularity.
As a requirement for a daily consensus buy signal of a stock, we set a threshold requiring that the number of submissions recommending an investment be 50\% higher than the number of submissions with a sell signal (and vice versa for a daily sell signal).

The resulting data is enriched with additional features in order to provide further flexibility during the analysis:
the number of all submissions posted on WSB, the average number of mentions over a window of three days, and the weekday. 
We also compute the relative change of the stock closing price \emph{since} and \emph{after} specific time windows: e.g., one day, one week, one month, three months.
Furthermore, we calculate the moving average of the closing price for seven, 30, and 90 days, and add a conditional buy signal which only counts WSB's and the professional investors' buy signals if the closing price is below the respective moving average.
We compile this data for each of the S\&P 500 stocks separately.

\subsection{Recommendations of Professional Investors}

\paragraphNoSpace{Data Compilation}
Investment recommendations of investment banks are usually based on extended analyst reports that are regularly published.
We compile a history of investment advice from financial institutions from the aforementioned Yahoo!\ Finance service. 
While the full reports are only available upon payment, the final verdicts, e.g., ``Buy'' or ``Hold'', are freely available. 
In order to select the professional investment banks to compare WSB to, we determine the top 20 institutions within the Yahoo!\ Finance data with regard to the number of investment recommendations they have made over the reviewed time frame.
While the number of published investment recommendations is not fully correlated with the banks' size and transaction volume, most of these institutions are nonetheless among the largest investment banks in the world.
The count of investment recommendations per institution is distributed as follows: Morgan Stanley (6,724), Credit Suisse (2,416), Citigroup (2,231), Wells Fargo (2,180), Barclays (1,987), Deutsche Bank (1,915), UBS (1,831), JP Morgan (1,507), Raymond James (1,503), BMO Capital (1,132), RBC Capital (1,007), KeyBanc (1,002), Goldman Sachs (859), Bank of America (843), BofA Securities (778), Mizuho (749), Baird (710), Jefferies (632), Piper Sandler (562),  Stifel Nicolaus (523).
These investment banks play an important role in the industry and are well-known and highly regarded for their analyses and decisions.
In light of this, we consider these 20 institutions as a suitable benchmark of financial professionals against which we can assess the performance of recommendation advice offered by WSB.

\paragraphWithSpace{Labels}
While there are similarities in how reports by different banks are created and published, there is still some variety in the exact wording of positive and negative investment recommendations among the different sources.
In order to standardize them, we summarized the 20 most common recommendation types into three different categories:
\begin{itemize}
    \item \textbf{Buy signals:} Contain all recommendations with ``Buy'', ``Overweight'', ``Outperform'', ``Strong Buy'', ``Positive'', ``Market Outperform'', ``Sector Outperform''.
    \item \textbf{Hold signals:} Contain all recommendations with ``Neutral'', ``Hold'', ``Equal-Weight'', ``Market Perform'', ``Sector Perform'', ``In-Line'', ``Sector-Weight'', ``Equal-weight'', ``Peer Perform''.
    \item \textbf{Sell signals:} Contain all recommendations with ``Underweight'', ``Underperform'', ``Sell''.
\end{itemize}

These three types of signals from each of the top 20 investment banks are aggregated on a daily basis per stock. 
In total, all investment recommendations of the top 20 professional investors from the Yahoo!\ Finance data sum up to 42,730 over the reviewed time frame of approximately 50 months. 
The distribution of these signals is highly skewed, as 24,097 represent a buy, 16,079 a hold, and only 2,554 a sell recommendation.
We attribute this to three main factors: 
During the reviewed time frame, the stock markets have shown an upward trend with the S\&P 500 gaining approximately 80\% in value (despite a temporary 30\% drop in March 2020 driven by the COVID-19 related stock market crash) up until the beginning of 2022, when stock prices started falling.
Additionally, it can be much more difficult to make accurate sell recommendations due to the nature of the stock market, which tends to increase in small increments in value over longer periods of time, while specific, unexpected events can cause a crash with significant value loss in a very short time window.
Finally, we are only reviewing S\&P 500 companies, which are well established and are expected to fulfill specific quality criteria that require a certain degree of financial success.

\section{Results}

\subsection{WSB and Institutional Portfolios}

\begin{figure*}[h]
    \centering
    \includegraphics[width=0.8\linewidth]{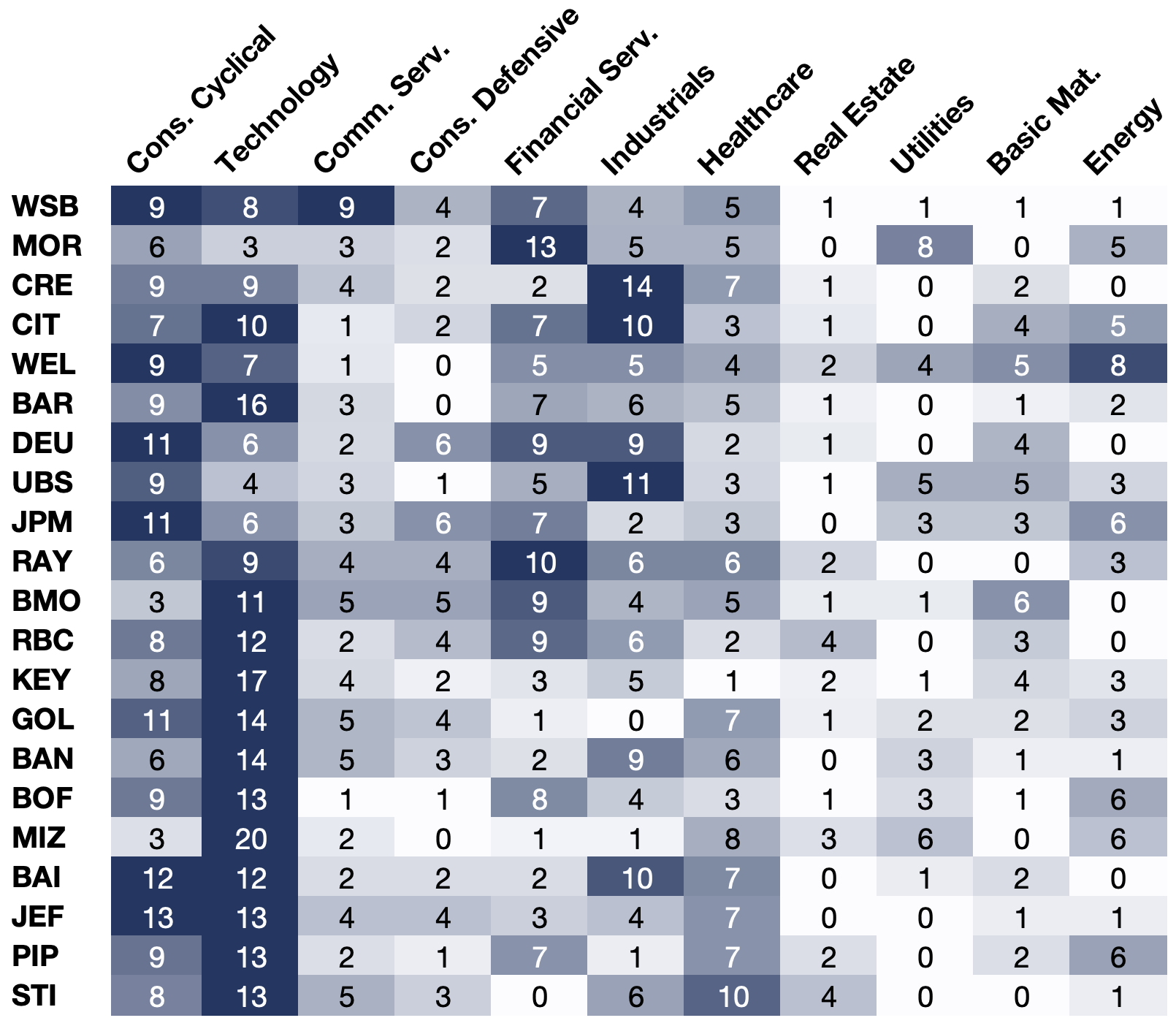}
    \caption{Distribution of investment recommendations per source (WSB and top 20 investment banks, abbreviated) across stock market sectors (coloured to indicate sectors of higher interest for each source)}
    \label{fig:portfolios}
\end{figure*}

First, for every source of investment advice (i.e., WSB and the top 20 investment banks), we define a portfolio consisting of their respective 50 most frequently recommended stocks.
Using these portfolios, we can compare how the choices of each investor are distributed across sectors.
Our results, given in Figure \ref{fig:portfolios}, show that most of the sources emphasize the sectors Consumer Cyclical and Technology, including Wells Fargo (WEL), Barclays (BAR), KeyBanc (KEY), Jefferies (JEF), and others.
We identify a second group including Morgan Stanley (MOR), Citigroup (CIT), UBS, and Deutsche Bank (DEU), which has a stronger focus on the sectors Financial Services and Industrials in addition to Consumer Cyclical, but less interest in the Technology sector. 
Additionally, some of the banks indicate higher interest in the Healthcare sector, e.g., Mizuho (MIZ), Stifel Nicolaus (STI). Selected few emphasize other sectors such as Utilities (Morgan Stanley) and Energy (Wells Fargo).
There are thus notable differences between the portfolios of similarly large institutions.
Potential reasons for this include a different strategic focus or varying domain expertise among their analysts.

While WSB's portfolio of S\&P 500 stocks exhibits a certain degree of similarity to those of some professional investors, the portfolio is distributed more evenly across a set of sectors, including Consumer Cyclical, Technology, Communication Services, and Financial Services.
We conclude from these insights that there is not a single standard investment strategy uniformly adhered to by all professionals, while WSB either hits or misses, but instead there are many different opinions and foci depending on each investor's inclinations and capabilities.
The fact that WSB's portfolio is actually fairly similar to multiple of the world's largest investment banks suggests that the community's focus may be less skewed towards a small part of the stock market than one might expect.

\subsection{Detecting Top-Performing S\&P 500 Stocks}

\paragraphNoSpace{Identification Criteria}
A common approach of evaluating the success of investments is to examine which stocks have been picked and how their values developed subsequently.
However, we first also consider a different perspective on this matter: Which S\&P 500 stocks have been the most successful and how many of them have actually been recommended by the different investors?

In order to answer this question, we consider two scores per stock: (1) the total change in value throughout the reviewed time frame 
(January 1, 2018 to March 22, 2022), and (2) the median three-month price development in percent.
For example, the 3M Company (\$MMM) was priced at 72.96\% of its initial value at the end of the reviewed time frame (\$205.52 in January 2018 and \$149.94 in March 2022), and exhibited a median price change of -0.37\% after three months.
While the former is a simple indicator for a stock's success over a time frame, the latter provides a better metric for how consistently the stock price has increased or decreased in value (stocks that have lost value most of the time, but then gained a large percentage at one point are unable to perform very well on the latter).
The top 15\% of the S\&P 500 stocks have been able to increase their stock price to at least 261.35\% of the initial value or have shown a median three-month price increase of 7.21\% (depending on which feature is chosen to select the top stocks).
We considered the stocks in the top 15\% with regard to their total growth as well as the stocks in the top 15\% with regard to the median three-month growth as the group of best-performing S\&P 500 stocks. 
This group consists of 56 stocks, including Apple, Tesla, Alphabet, Nvidia, Moderna, MSCI, AMD, and Microsoft, among others.

\begin{figure*}[ht]
    \centering
    \includegraphics[width=0.8\linewidth]{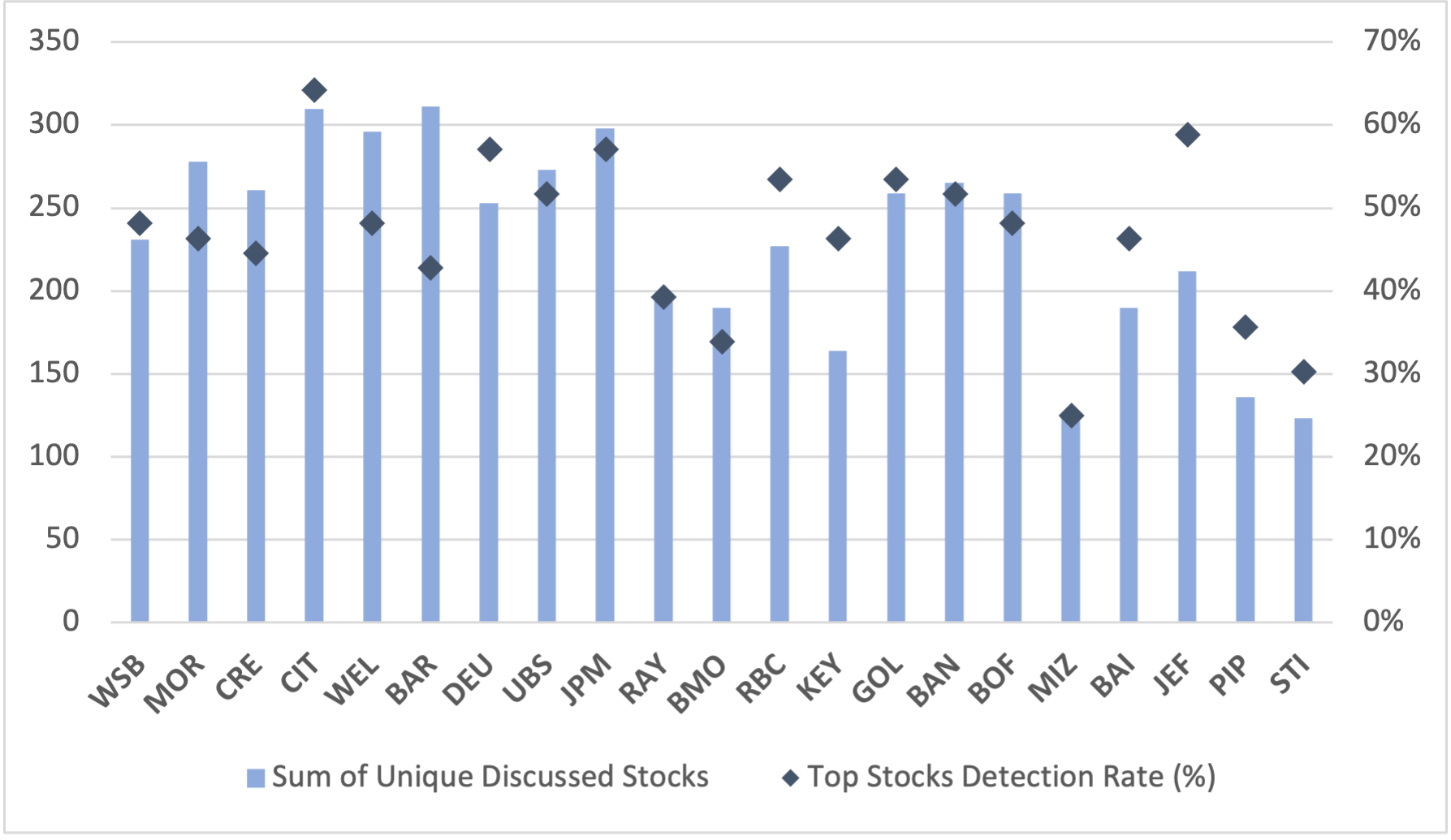}
    \caption{Unique Discussed Stocks and Detection Rate of Best Performing Stocks per Investment Signal Source (WSB and Investment Banks)}
    \label{fig:stocks_detection}
\end{figure*}

\paragraphWithSpace{Comparison of Recommendations}
Having identified S\&P 500's top-performing stocks, we compared this list with the stocks that WSB and each of the financial institutions recommended over the same time frame.
At this point, it should be noted that while the number of buy signals published by the top 20 professional investors ranges from 285 to 2,693 per investor, our methodology for detecting signals in WSB introduced earlier has detected 9,868 buy signals on WSB.
Clearly, this is a substantial difference, as WSB is a community of millions of users voicing their opinions and analyses. 
However, these WSB recommendations are much more repetitive, and ultimately still amount to a very similar number of unique recommended companies:
Overall, the large number of considered WSB recommendation signals still just correspond to a set of 231 unique companies, with some of the investment banks coming close to or surpassing this number, e.g., Morgan Stanley (278), Credit Suisse (261), Citigroup (310), Wells Fargo (296), Barclays (311), and UBS (273).
Thus, despite the systematic difference of how many voices publish recommendations (millions of users on WSB versus a single institution per investment bank), the number of unique companies that the investors have recommended is fairly comparable (see Figure \ref{fig:stocks_detection}).

We can thus proceed to answer the question.
Our analysis shows that WSB has performed well compared to the investment banks regarding the detection rate of top performing S\&P 500 stocks:
WSB has detected 27 of the 56 top performing stocks, with seven banks reaching a higher ratio: Citigroup (36), Deutsche Bank (32), JP Morgan (32), RBC Capital (30), Goldman Sachs (30), Bank of America (29), Jefferies (33). Two banks reached the same ratio: Wells Fargo (27), B of A Securities (27). 
While there are multiple banks reaching a higher ratio of detected top stocks, it has to be noted that all except for RBC Capital and Jefferies have a higher number of unique discussed stocks and therefore a higher chance to mention the right stocks.
The results are similar when applying the same method to all S\&P 500 stocks that only fulfill one of the two criteria (top 15\% in total growth or top 15\% in median three-month growth).

\paragraphWithSpace{Discussion}
We conclude from this analysis that not only has the WSB community performed well in detecting the right companies and recommending an investment in them, but for the considered time period they also produced a better selection than a number of investment banks with teams of analysts working on this.
While investment banks require their customers to pay substantial fees to obtain access to their analyses, the recommendations of WSB are freely accessible and even include the reactions and opinions of other community members to each recommendation. 
However, investment banks tend to provide their analysis in a convenient condensed form, while discerning valuable advice among the numerous posts on WSB requires more effort.
For a real WSB user to adopt our data-driven approach, they would have to spend a sufficient amount of time visiting the community on a regular basis to recognize the most actively discussed and recommended stocks within a time window such as 24 hours to be able to infer the community consensus on all relevant stocks.

\subsection{Evaluation of Buy Signals}

The next part of our study assesses more specifically what performance the specific buy recommendations encountered on WSB achieved in comparison to buy recommendations made by professional investors. 
For this evaluation, we focus on \emph{buy signals}, as our prior analysis has shown that sell signals performed poorly due to the general upwards trend of the stock markets (except in a few cases in which sell signals achieved short-term success). 
Hold signals provide limited value for investors looking to make new investments and are therefore more difficult to include in an investment strategy.
In order to evaluate the buy signals from the data described in the previous section, we create an aggregated overview of all of the 500 reviewed stocks.
We define two main metrics for evaluating a buy signal's success: the accuracy and the price performance.
Our first metric, the accuracy of buy signals, measures the percentage of signals that have experienced an increase in value at all.
The second metric, the price performance, measures the average price change after a hypothetical investment.
For both metrics, we review the stock price on exact time windows after each buy signal, specifically 
one week, one month, and three months.

\begin{table} 
  \centering
  \caption{Average accuracy and price performance of buy signals from WSB and selected investment banks (evaluated one week, one month, and three months after signal; ``WSB (prox)'' indicating WSB buy signals detected in proximity of tickers;  ``MA'' indicating the moving average condition)}
  \label{tab:buy_acc_perf} 
  \begin{tabular}{lcccccc}
                                        & \multicolumn{3}{c}{Accuracy after} & \multicolumn{3}{c}{Price Change (\%) after} \\
                 Source                 &  1 w   & 1 m   & 3 m  &  1 w   & 1 m   & 3 m  \\
    \midrule
                 WSB                    &  0.551    &  0.600    &  0.699    &  0.532    &  2.240    &  6.947    \\
                 WSB (prox)             &  0.500    &  0.500    &  0.667    &  0.090    &  0.695    &  4.823    \\
                 WSB (MA30)             &  0.545    &  0.667    &  0.750    &  0.629    &  3.049    &  9.237   \\
                 WSB (MA90)             &  0.555    &  0.667    &  0.800    &  0.690    &  6.325    &  9.565   \\
                 Morgan Stanley         &  0.538    &  0.600    &  0.615    &  0.355    &  2.030    &  4.754    \\
                 Credit Suisse          &  0.500    &  0.588    &  0.600    &  0.579    &  1.603    &  3.907    \\
                 Citigroup              &  0.500    &  0.625    &  0.636    &  0.373    &  1.627    &  4.290    \\
                 Wells Fargo            &  0.600    &  0.600    &  0.600    &  0.907    &  1.910    &  3.785    \\
                 Barclays               &  0.500    &  0.500    &  0.500    &  0.121    &  1.153    &  3.450    \\
                 Deutsche B.            &  0.545    &  0.539    &  0.667    &  0.472    &  1.452    &  5.614    \\
                 UBS                    &  0.500    &  0.600    &  0.667    &  -0.058   &  1.577    &  5.524    \\
                 JP Morgan              &  0.600    &  0.667    &  0.500    &  0.616    &  2.152    &  2.717    \\
                 Raymond James          &  0.556    &  0.583    &  0.600    &  0.554    &  1.408    &  4.842    \\
                 BMO Capital            &  0.500    &  0.600    &  0.600    &  0.487    &  1.710    &  3.088    \\
                 RBC Capital            &  0.600    &  0.667    &  0.667    &  0.674    &  1.559    &  3.754    \\
                 KeyBanc                &  0.538    &  0.667    &  0.714    &  0.355    &  2.471    &  5.600    \\
                 Goldman Sachs          &  0.500    &  0.667    &  1.000    &  0.347    &  2.155    &  3.717    \\
                 Bank of America        &  0.667    &  0.667    &  0.500    &  0.653    &  1.163    &  1.013    \\
                 B Of A Securities      &  0.667    &  0.667    &  1.000    &  0.991    &  1.347    &  8.442    \\
                 Mizuho                 &  0.625    &  0.500    &  0.500    &  0.807    &  0.712    &  4.426    \\
                 Baird                  &  0.500    &  0.500    &  0.667    &  -0.138   &  -0.158   &  4.728    \\
                 Jefferies              &  0.500    &  0.667    &  0.667    &  0.347    &  1.598    &  3.135    \\
                 Piper Sandler          &  0.500    &  0.500    &  0.600    &  0.394    &  -0.006   &  4.908    \\
                 Stifel Nicolaus        &  0.500    &  0.500    &  0.500    &  -0.143   &  0.271    &  1.273    \\
                 
    \bottomrule
  \end{tabular}
\end{table}

\paragraphWithSpace{Accuracy of Buy Signals}
An analysis of the accuracy of buy signals shows that the WSB community is able to attain the level of performance of the top investment banks -- the results are better than most of the top 20 investment banks.
While the short-term performance is quite similar among almost all sources and close to 50\%, the differences in accuracy become clearer when looking at the average accuracy after the three-month time window.
Table~\ref{tab:buy_acc_perf} shows that approximately 70\% of WSB buy signals (for S\&P 500 stocks) would have led to a positive price development after three months, placing WSB close to the best of the top 20 investment banks, of which only three have achieved better accuracy in the three-month time window.
Interestingly, 100\% of buy signals from Goldman Sachs and B Of A Securities have increased in price after three months.
WSB's accuracy can be further improved by filtering for the buy signals that occurred on days when the stock price was below the calculated moving averages of the last 30 or 90 days, with the best accuracy being 80\% when utilizing the moving average of the last 90 days.

While this metric does not reflect how well the buy signals have worked in terms of a relative price increase, it is an indicator of how consistently a buy recommendation could have been trusted.
Hence, if an investor's main concern is reliability and positive price development even if they do not receive the highest yields, this analysis approach could help in evaluating buy recommendations.
It should be noted that due to the sizes and role as market leaders of these large investment banks, the accuracy of their recommendations could also be positively influenced by investors following their recommendations, while the respective stock might otherwise not have received the same amount of attention.
Pursuing this question is non-trivial and beyond the scope of this paper.

\paragraphWithSpace{Price Performance of Investment Signals}
In our second metric, the price performance, WSB's performance is even more competitive, beating most of the investment banks.
When filtering WSB buy signals using the moving average condition, the average price increase in fact significantly outperforms all investment banks.
Our choice of the moving average as an additional condition is due to its ability to serve as a simple method of confirming whether a stock has recently experienced increases in its stock price -- if this is the case, a buy recommendation may be too late and therefore reactive.
As evinced in Table \ref{tab:buy_acc_perf}, WSB's buy signals achieve an average price increase of 6.947\% after three months and over 9.2\% and 9.5\% when filtered with a moving average of 30 and 90 days, respectively.
In comparison, the best investment banks achieve 8.44\% (BofA Securities), 5.61\% (Deutsche Bank), 5.60\% (KeyBanc), 5.52\% (UBS).
This means that an investor following all of WSB's buy signals over the reviewed time frame would have achieved similarly high profits as one that followed the most successful investment banks, when selling after three months.
An investor that followed only WSB signals fulfilling the moving average condition would have achieved substantially higher profits.

\paragraphWithSpace{Discussion}
These results suggest that following WSB's buy recommendations appears to be riskier, as not all of them lead to a positive price development after three months, while investment banks to a larger extent recommended stocks that actually increased in price. 
On average, however, investment advice from WSB yielded higher profits, leading to a better financial outcome on average, even if some of the buy signals turned out to be wrong.
These results only account for WSB's buy signals of S\&P 500 stocks, which excludes many of the frequently discussed meme stocks -- however, this does not seem to be a disadvantage for WSB.
On the contrary, filtering WSB's recommendations for more established companies could even be beneficial, as they are rarely as volatile as the stocks of smaller companies, reducing the chances of rapid price increases followed by sudden crashes, which often lead to financial losses for slower or less experienced investors.

The reader should be aware that a large portion of the data covers a phase during which the stock markets have experienced an upwards trend (with some exceptions, e.g., interest rate changes in 2018 or the Covid-related crash in 2020). 
In 2022, the market has changed significantly due to multiple factors including the war in Ukraine, rising inflation, interest rate hikes, leading to a so-called ``bear market'' with large losses in some stock prices and markets.
Once more time has passed and there is more data available of markets in a downward trend, we are planning to extend our analysis for the changed situation in order to reveal whether this is a temporary crash similar to 2020 or a more permanent situation.
Our initial analysis indicates that while the WSB cannot avoid losses in 2022, these seem to be slightly more moderate than the S\&P 500's: While the S\&P 500 has changed by -13.06\% on average after three months in the first quarter of 2022, the WSB signal baseline indicates a change of -12.00\%.
In future work, we aim to extract more value from WSB's investment signals by developing a machine-learning-based methodology to identify signals to trust and those to ignore.

\subsection{WSB After the Hype}

As explained in the first sections of this paper, the WallStreetBets community witnessed substantial growth due to the GameStop hype in January and February 2021: The subscriber count quintupled and even international mainstream news reported on the phenomenon.
Clearly, an event of this magnitude has the potential to influence and alter the make-up of a community in  multiple respects.
In the context of this paper, the most significant question is whether the hype and user growth affected the community's ability to make valuable investment recommendations.
In a recent update to their paper, Bradley et al.~\cite{bradley2021place} argue that after the GameStop hype, the quality of investment advice from WSB appears to have deteriorated.
The authors hypothesize that after (seemingly) successfully affecting the stock price of GameStop, the community may have tried to repeat the success by initiating new coordinated trading strategies. 
They conclude that recommendations on WSB should be filtered for quality before placing any trust in them, which is also in line with our study. 
During our analysis of the dataset, we observed an increase in stocks that can be considered meme stocks (often of smaller, less established companies) -- possibly fuelled by the significant increase in WSB users and their thirst for another short squeeze like GameStop's.

In order to pursue this question, we repeated the analysis from the previous section on data from 2018 to 2020 (pre-hype) and 2021 (post-hype) separately.
When viewing WSB recommendations of S\&P 500 companies, our analysis shows some differences between the pre- and post-hype performance, but not a complete reversal:
Regarding the accuracy of buy signals, WSB does seem to have degraded in quality quite noticeably, as pre-hype buy signals achieved 79\% accuracy for all buy signals and 100\% for those with a moving average 30 or 90 condition (MA30/90), while post-hype signals achieved 55\% and 67\%, respectively.
In the same comparison, the investment banks have maintained similar levels of accuracy or even improved in the post-hype time frame.
There is a pre- and post-hype difference in the price performance as well: in the pre-hype period, WSB achieves an 8.2\% average price change after three months for all buy signals and approximately 11\% for buy signals with the MA30/90 condition, which decreases to 7.3\% and approximately 9\%, respectively, when only the post-hype period is considered.
However, the investment banks also achieved lower average profits in the post-hype time frame, which enables WSB to perform similarly to the best investment bank (BMO Capital, 9.15\%), continuing to rank close to the top of the comparison (followed by KeyBanc at 8.2\%, Credit Suisse at 7.15\%, and UBS at 7.14\%).
 
We conclude from our analysis that while there are some differences between the pre- and post-hype states of WSB, they are less visible when focusing on stocks of the established companies listed on the S\&P 500.
The lower price performance in 2021 apparently affected all investors and might have been due to larger market effects, as WSB continues to achieve a leading price performance in the post-hype time frame.

\section{Conclusion}

Our results show that the community of WSB not only competes with, but in some cases even beats the returns of analyst reports published by top 20 investment banks when comparing investment recommendations on S\&P 500 stocks.
WSB is able to recognize high-potential stocks better than professional analysts, and achieve similarly high average returns as the best of the reviewed investment banks.
When WSB's signals are filtered using a moving average condition, the average returns are significantly higher than those of banks.

Our analysis indicates that WSB as an openly accessible social community can be a valuable source of information for retail and professional investors due to its open collaborative process of discussing stock market trends and recommending investments.
However, it is important to filter WSB's wealth of signals using suitable techniques, e.g., focusing on more established companies like those of the S\&P 500 index, or filtering signals using simple indicators.
Actively discussed meme stocks should be handled with care due to their higher risk and the difficulty of identifying the right time to invest in them.
Since the GameStop hype, meme stocks have been appearing more frequently on WSB, which makes filtering more difficult as well as important.
If social communities like WSB continue to grow and manage to prove the quality of their content, this may profoundly change the way investors seek financial advice, especially retail investors, who are on the rise due to the increasing popularity of mobile trading applications.
Professional investors, while presumably being well informed about the stock markets already, can leverage social communities like WSB to achieve a better understanding of retail traders, who constitute an increasingly important group that is sometimes able to exert notable influence on the stock markets.

\newpage
\section*{Additional Information}

\subsection{Dataset Details}

Table \ref{tab:reddit_features} shows examples for the features of the Reddit dataset extracted via the Pushshift API.

\begin{table}[h]
  \centering
  \caption{Example features from the Reddit dataset}
  \label{tab:reddit_features}
  \begin{tabular}{lp{6cm}}
    \toprule
    Category & Examples for features of Reddit posts \\
    \midrule
    Author      & author, author\_flair\_type, author\_flair\_text, author\_is\_blocked \\
    Details     & id, is\_original\_content, is\_video, media\_only, over\_18, spoiler \\
    Design      & author\_flair\_css\_class, link\_flair\_text\_color, link\_flair\_background\_color \\
    Metadata    & awarders, is\_crosspostable, is\_meta, pinned, removed\_by\_category \\
    Links       & domain, full\_link, permalink, url \\
    Performance & all\_awardings, gildings, num\_comments, num\_crossposts, score, total\_awards\_received, upvote\_ratio \\
    Subreddit   & subreddit\_id, subreddit\_subscribers, subreddit\_type \\
    Text        & title, selftext, link\_flair\_text \\
    Timestamp   & created\_utc, retrieved\_on \\
  \bottomrule
\end{tabular}
\end{table}

\bibliographystyle{unsrt}  
\bibliography{references}

\end{document}